\documentclass[twocolumn,revtex4-2]{openjournal} 
\usepackage[T1]{fontenc}
\usepackage{ae,aecompl}
\usepackage[dvipsnames]{xcolor}
\usepackage{graphicx}	
\usepackage{amsmath}	
\usepackage{amssymb}	

\usepackage{natbib}

\usepackage{hyperref}
\hypersetup{
	colorlinks=true,
	urlcolor=blue,    
	linkcolor=black,  
	citecolor=blue,   
}

\defcitealias{Pittordis_2018}{PS18}
\defcitealias{Pittordis_2019}{PS19}
\defcitealias{Badry_2021}{ERH}
\defcitealias{Banik_2018_Centauri}{BZ18}

\hypersetup{citecolor=blue, 
            linkcolor=red, 
            menucolor=blue, 
            urlcolor=blue}  

\newcommand{\cm}{{~\rm cm}}

\newcommand{\km}{{~\rm km}}
\newcommand{\s}{{~\rm s}}

\newcommand{\erg}{{~\rm erg}}
\newcommand{\yr}{{~\rm yr}}

\newcommand{\AU}{{~\rm AU}}

\begin{document}

\title{The compact circumstellar material of SN~2024ggi: Another supernova with a pre-explosion effervescent zone and jet-driven explosion}


\author{Noam Soker}
\affiliation{Department of Physics, Technion, Haifa, 3200003, Israel; soker@technion.ac.il}


\begin{abstract}
I examine recent theoretical studies and observations of the recent core-collapse supernova (CCSN) SN~20224ggi and find that the likely explanation for its dense, compact circumstellar material is an effervescent model, where parcels or streams of gas are uplifted by stellar convection and pulsation and fall back. The effervescent zone 
exists alongside the regular wind from the red supergiant (RSG) progenitor of SN~2024ggi. I find that an extended wind-acceleration zone encounters some difficulties in accounting for the required CSM mass. Recent modelling finds the explosion energy of SN~2024ggi to be $E_{\rm exp} > 10^{51} \erg$, and up to $E_{\rm exp} \simeq 2 \times 10^{51} \erg$. I examine this explosion energy against a recent study of the delayed neutrino explosion mechanism and find that this mechanism  might have some difficulties in accounting for the required energy. This  might suggest that the explosion was caused by the jittering jets explosion mechanism (JJEM). This adds to other recent pieces of evidence supporting the JJEM, particularly point-symmetric CCSN remnants.    
\end{abstract}


\section{Introduction}
\label{sec:intro}

Studies fit the type IIP core-collapse supernova (CCSN) SN~2024ggi lightcurve and spectroscopic properties with a compact pre-explosion circumstellar material (CSM; e.g., \citealt{ChenTW2024, ChenXetal2024, JacobsonGalanetal2024, Pessietal2024, Shresthaetal2024, XiangD2024, ZhangJ2024}). By compact CSM, I refer to CSM with which the ejecta collides within days after the explosion, unlike SN~1987A, where a collision occurs years after the explosion. In principle, compact dense CSM of CCSN progenitors might result from an enhanced mass loss rate that starts years to weeks before explosion, possibly accompanied by a pre-explosion outburst (e.g., \citealt{Foleyetal2007, Pastorelloetal2007, Smithetal2010, Marguttietal2014, Ofeketal2014, SvirskiNakar2014, Tartagliaetal2016, Yaronetal2017, Wangetal2019, Bruchetal2020, Prenticeetal2020, Strotjohannetal2021, JacobsonGalanetal2022}), an extended long-lived dense zone above the stellar photosphere (e.g., \citealt{Dessartetal2017}), an extended accelerated zone of the wind (e.g., \citealt{Moriyaetal2017, Moriyaetal2018}) or by a long-lived extended dense zone of uprising and falling gas parcels or streams above the stellar photosphere (e.g., \citealt{Soker2021effer, Soker2023Effer, FullerTsuna2024}; see further discussion by \citealt{FullerTsuna2024}). SN~2024ggi did not experience an outburst within years before the explosion (e.g., \citealt{Shresthaetal2024}).

The properties of the compact CSM of SN~2024ggi are not much different from those of SN~2023ixf (for SN~2023ixf, e.g., \citealt{Bergeretal2023,  Bostroemetal2023, Grefenstetteetal2023, JacobsonGalanetal2023, Kilpatricketal2023, SinghTejaetal2023, SmithNetal2023}), which also did not have any pre-explosion outburst that could have formed the compact CSM (e.g., \citealt{Jencsonetal2023, Neustadtetal2023, Soraisametal2023}). In \cite{Soker2023Effer}, I argued that the best explanation for the compact pre-explosion CSM of SN~2023ixf is an effervescent zone. In this study, I make a similar claim for the compact CSM of SN~2024ggi. 

Most studies of SN~2024ggi mentioned above ignore the effervescent zone model and the similar model by \cite{FullerTsuna2024}, which includes many more details; in both models, convective motion and pulsation uplift the bound material, but \cite{FullerTsuna2024} calculate the uplifting of material by shocks that the convection excites above the photosphere. In this study, I use the term effervescent zone model to include both the model studied by \cite{Soker2023Effer} and \cite{FullerTsuna2024}. In some aspects, but not all, the hydrostatic model of \citealt{Dessartetal2017} can also be grouped with the effervescent zone model. 

In section \ref{sec:CSM}, I examine the proposed wind acceleration zone that \cite{ChenTW2024} construct for the dense compact CSM of SN~2024ggi. I reiterate my claim that the effervescent zone should not be ignored. 
In section \ref{sec:Eexp} I comment on the interesting finding by \cite{ChenTW2024} that their best fit gives an explosion energy of $E_{\rm exp} \simeq 2 \times 10^{51} \erg$. This has implications for the likely explosion mechanism of SN~2024ggi that I emphasize. In my summary in section \ref{sec:Summary}, I comment on the relation between the effervescent zone model and the explosion mechanism. 

\section{The compact CSM of SN~2024ggi}
\label{sec:CSM}

A most recent study and analysis of the compact CSM of SN~2024ggi is that by \cite{ChenTW2024}. They consider the wind acceleration model of \cite{Moriyaetal2018} for a compact CSM that extends to $R_{\rm CSM} \simeq 6 \times 10^{14} \cm$, and that has a mass loss rate of $\dot M_{\rm CSM} \simeq 0.001 M_\odot \yr^{-1}$ for a wind terminal velocity of $v_{\rm w} = 10 \km s^{-1}$. The wind velocity is lower near the star as it is accelerated, such that the total wind mass within the radius of $R_{\rm CSM}$ is $M_{\rm CSM}  \simeq 0.4 M_\odot$. 
There are two challenges to this CSM model, as I discuss next. 

The first challenge is to explain what accelerates the wind. The reason is that the momentum in the radiation is much smaller than that of the wind.
Momentum balance limits the mass loss of the wind that radiation can accelerate. It is given by the equation  
$\dot M_{\rm wc} v_{\rm w} \simeq \eta_{\rm w} L/c$, where $v_{\rm w}$ is the terminal wind speed, $L$ the stellar luminosity, and $\eta_{\rm w}$ is the average number of times that a photon transfers its momentum to the wind in the outward radial direction,  
\begin{eqnarray}
\begin{aligned} 
\dot M_{\rm wc} & \simeq  4 \times 10^{-4} \eta_{\rm w}
\left( \frac{L}{2 \times 10^5 L_\odot} \right)
\\ & \times 
\left( \frac{v_{\rm w}}{10 \km \s^{-1}} \right)^{-1}
M_\odot \yr^{-1} .
\label{eq:mwc2}
\end{aligned}
\end{eqnarray}
The mass loss rate of the wind-accelerated zone of \cite{ChenTW2024} requires $\eta_{\rm w} \simeq 2.5$. This requires several scattering of the photons, which is unclear how to achieve to distances of $>10 \AU$ from the star. 
With the same model \cite{Singhetal2024} estimate the mass loss rate to build the CSM of SN~2023ixf to be $0.001-0.01 M_\odot \yr^{-1}$, which requires the highly-unlikely value of up to $\eta_{\rm w}\simeq 25$.

The demand for $\eta_{\rm w} \simeq 2.5$ at distances of tens of AU from the star is challenging for the wind-acceleration zone model. Note that the number $\eta_{\rm w}$ does not refer to the total scattering of photons but to the scattering in the radial direction, such that the photon deposits a radial outward momentum. A simple optically-thick medium does not increase $\eta_{\rm w}$ above 1. One needs scattering from different sides of the star, namely, that photons bounce from one side to another.  

The terminal wind velocity that \cite{ChenTW2024} and \cite{MoriyaSingh2024} take in their model, $v_{\rm w} = 10 \km \s^{-1}$, is much lower than the escape velocity from RSG progenitors of SNe II,  e.g., $v_{\rm escape} = 39 \km \s^{-1}$ for an RSG with a mass and radius of $M_{\rm RSG}=4 M_\odot$ and $R_{\rm RSG}=1000 R_\odot$, respectively. 
From the high-resolution spectra of SN~2024ggi,  \cite{Shresthaetal2024} calculated a CSM velocity of $37 \pm 4 \km \s^{-1}$.  
The low terminal wind velocity that \cite{ChenTW2024} take is even more problematic as this is lower than the escape velocity at the outer boundary of the compact dense CSM 
\begin{eqnarray}
\begin{aligned} 
v_{\rm w} & = 10 \km \s^{-1} < v_{\rm esc} (R_{\rm CSM}) =  13.3
\\ & \times
\left( \frac{M_{\rm RSG}}{4 M_\odot} \right)^{1/2} 
\left( \frac{R_{\rm CSM}}{6 \times 10^{14} \cm} \right)^{-1/2} \km \s^{-1}.
\label{eq:Vesc}
\end{aligned}
\end{eqnarray}
The RSG mass is likely higher than the above scaling of $M_{\rm RSG} = 4 M_\odot$. This implies that the acceleration zone must continue beyond $R_{\rm CSM} = 6 \times 10^{14} \cm$. 
 Note that the progenitor cannot have a mass much below $M_{\rm RSG}=4 M_\odot$, as a mass of $\simeq 1.5 M_\odot$ is in the neutron star, and the ejecta should contain few solar masses or more. 

The most important and robust parameter of the CSM that \cite{ChenTW2024} and \cite{MoriyaSingh2024} obtain in their modeling is the CSM mass. If the terminal wind velocity is taken to be higher, as should be for such RSG stars, by a factor of $k_{\rm w}$, then the mass loss rate should also be $k_{\rm w}$ higher. This implies a wind radial momentum larger by a factor of $k^2_{\rm w}$, and so is the value of $\eta_{\rm w}$. This would make the value of $\eta_{\rm w} \gg 10$ unrealistically high.

This section concludes that the wind-accelerated zone model for a compact and dense CSM of SN~2023ixf and SN~2024ggi encounters severe challenges. I consider this model unlikely to explain the CSM of SN~2024ggi and SN~2023ixf. However, the total CSM mass and the explosion energy that \cite{ChenTW2024}, \cite{Singhetal2024}, and \cite{MoriyaSingh2024}  obtain in their modeling hold.   

\section{The implication of $E_{\rm exp} \ga 2 \times 10^{51} \erg$}
\label{sec:Eexp}

With their accelerated-wind zone model for the CSM, \cite{ChenTW2024} estimated the explosion energy of SN~2024ggi to be $E_{\rm exp} \simeq 2 \times 10^{51} \erg$, and \cite{Singhetal2024} (see also \cite{MoriyaSingh2024}) estimate the explosion energy of SN~2023ixf to be $E_{\rm exp} \simeq 2-3 \times 10^{51} \erg$. Their fitting might apply to the effervescent zone model if it has a similar density profile. \cite{JacobsonGalanetal2024} adopted an explosion energy of $E_{\rm exp} \simeq 1.2 \times 10^{51} \erg$ for SN~2024ggi. Such explosion energies challenge the delayed neutrino explosion mechanism of CCSNe, which cannot account for explosion energies of $E_{\rm exp} \ga 2 \times 10^{51} \erg$, and even for $E_{\rm exp} \ga 10^{51} \erg$ it has severe difficulties (e.g., \citealt{Fryeretal2012, Ertletal2016, Sukhboldetal2016, Gogilashvilietal2021}). 
To emphasize this limitation of the delayed neutrino explosion mechanism, I critically examine a most recent study by \cite{Nakamuraetal2024}, who write, ``We find that neutrino-driven explosions occur for all models within 0.3 s after bounce.'' 

\begin{figure*}
\begin{center}
\includegraphics[trim=0.0cm 18.5cm 4.0cm 0.0cm ,clip, scale=1.05]{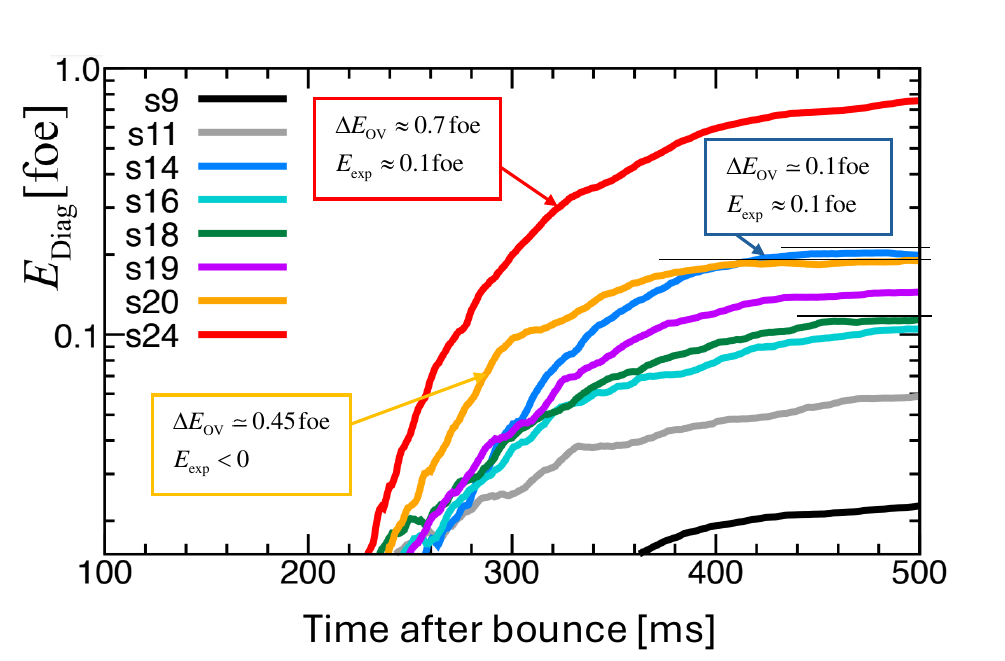} 
\end{center}
\caption{A figure adapted from \cite{Nakamuraetal2024}. The different lines show the evolution of the diagnostic energy in their simulations of the delayed neutrino explosion mechanism of stellar models differ in their zero-age main sequence mass, as indicated on the inset on the left (foe stands for fifty-one erg, i.e., $10^{51} \erg$). The diagnostic energy includes only grid cells with positive energy that moves radially outwards. With three horizontal black lines, I emphasize cases where the diagnostic energy does not increase at the end of their simulation, or, for $M_{\rm ZAMS} = 14 M_\odot$, even decreases at the end. In the inset attached to the three cases with the largest diagnostic energies, I write the estimated binding energy of the stellar parts with negative energy, namely the binding energy $\Delta E_{\rm OV}$ (overburden; based on \citealt{Mezzacappaetal2015}), as well as the expected final explosion energy $E_{\rm exp}=E_{\rm Diag}-\Delta E_{\rm OV}$: 
$E_{\rm exp}(14 M_\odot)\simeq 0.1 \times 10^{51} \erg$ and 
$E_{\rm exp}(24 M_\odot)\simeq 0.1 \times 10^{51} \erg$. For the cases of 
$M_{\rm ZAMS} = 16, 18, 19, 20$, the overburden is larger than the diagnostic energy, and I expect no explosion (or one with very low energy of $\ll 0.1$~foe) according to the simulations by \cite{Nakamuraetal2024}. I mark this only for  $E_{\rm exp}(20 M_\odot)< 0$. 
}
\label{Fig:Figure1}
\end{figure*}

\cite{Nakamuraetal2024} conduct three-dimensional magnetohydrodynamical simulations of the delayed neutrino explosion mechanism for stellar models with different zero-age main sequence masses from $9 M_\odot$ to $24 M_\odot$. Their stellar models are from \cite{Sukhboldetal2016}. The diagnostic explosion energy of their results is the sum of all numerical cells where the total energy is positive, and the materials have a positive radial velocity. The integrand is the sum of kinetic, gravitational, internal, and magnetic energy. In Figure \ref{Fig:Figure1}, I present their figure of the evolution of the diagnostic explosion energy, $E_{\rm Diag}$ with time. 

The final explosion energy is the diagnostic energy minus the binding energy of the rest of the envelope and core (termed overburden), 
namely, $R_{\rm exp}=E_{\rm Diag}-\Delta E_{\rm OV}$. Because \cite{Nakamuraetal2024} do not give the binding energy of the rest of their stellar models (overburden), I take overburden energies from the simulations of \cite{Mezzacappaetal2015} who give this for 4 models at $t \simeq 1.2 \s$ when the shock radius is larger than those at the end of the simulations by \cite{Nakamuraetal2024} each at $t=0.3 \s$. This implies that the binding energy (overburden) values should be larger even. 
 In calculating the overburden energy, I use the graphs from \cite{Mezzacappaetal2015} that already include the recombination energy of free nucleons; in any case, this energy is small. 
Also, \cite{Mezzacappaetal2015} and \cite{Nakamuraetal2024} do not use the same models. Therefore, the binding energy I give in the insets in Figure \ref{Fig:Figure1} are approximate values. These values, nonetheless, clearly point to the very low final explosion energies (to distinguish from the diagnostic explosion energy) that \cite{Nakamuraetal2024} obtain. 
Those models that explode have energies of $E_{\rm exp} \lesssim 0.1 \times 10^{51} \erg$ while some others seem to have negative energy, namely, they do not explode, e.g., the simulations for $M_{\rm ZAMS} = 16, 18, 19, 20$ lead to no explosion.

For the present study, relevant is the limit of $E_{\rm exp} < 1.2 \times 10^{51} \erg$ of most simulation results of the delayed neutrino explosion mechanism. In their very long simulation to 7 seconds after bounce, \cite{Bolligetal2021} find the explosion energy of a model with $M_{\rm ZAMS} = 19 M_\odot$ to be $\simeq 10^{51} \erg$. Even the diagnostic explosion energies of \cite{Nakamuraetal2024} do not reach these values. \cite{Mezzacappaetal2015} obtain higher explosion energies than \cite{Nakamuraetal2024}, but still of $E_{\rm exp} < 10^{51} \erg$. \cite{Burrowsetal2024} obtain higher explosion energies, but still most models have $E_{\rm exp} < 1.2 \times 10^{51} \erg$. Their results do not agree with those of \cite{Nakamuraetal2024}, e.g., \cite{Burrowsetal2024} obtain clear explosion with $E_{\rm exp} \simeq 10^{51} \erg$ in models with negative value of $E_{\rm exp}$ in the simulations of \cite{Nakamuraetal2024}, e.g., $M_{\rm ZAMS} = 16, 18, 19, 20$. Such qualitative disagreements between groups simulating the delayed neutrino explosion mechanism are a severe problem for this model. 

The conclusion from the short discussion above is that the modelings of SN2024ggi and SN2023ixf suggest that the  delayed neutrino mechanism might encounter some difficulties in explaining their explosion energy. 
Adding to this energetic consideration the challenge of the delayed neutrino mechanism to explain the point-symmetric morphologies of some CCSN remnants identified in the last two years (e.g.,  \citealt{BearSoker2024Cass, Soker2024NewA1987A, Soker2024Rev, Soker2024CFs}), I conclude  with my view that the likely explosion mechanism of the progenitors of SN~2024ggi and SN~2023ixf was the jittering jets explosion mechanism (JJEM).
 In the JJEM, several to few tens of pairs of opposite jets with fully or partially stochastic varying directions explode the star. Neutrino heating exists but does not play the primary role in the explosion process. Intermittent accretion disks around the newly born neutron star launch the pairs of jets over $\simeq 0.5-10 \s$; one to few later pairs of jets are also possible (for more quantitative values of the JJEM see \citealt{Soker2024Keyhole}). The source of the stochastic angular momentum of the gas that the newly born neutron star accretes from the collapsing core is the convective motion in the silicon and oxygen-burning zones of the pre-collapse core. Instabilities above the neutron star and below the shock of the collapsing core at $\simeq 150 \km$ from the center amplify the angular momentum fluctuations of the convective motion. Such are the spiral modes of the standing accretion shock instability (for a recent study of this instability, see, e.g.,  \citealt{Buelletetal2023}).
The JJEM can also account for the polarization of CCSNe and its correlation with the explosion energy (for the correlation, see  \citealt{Nagaoetal2024b}), as well as for bipolar CCSNe (e.g., \citealt{Nagaoetal2024a}). 

\section{Summary}
\label{sec:Summary}

The two recent relatively near CCSNe SN~2023ixf and SN~2024ggi allow good fitting to their light curves and spectra, suggesting a dense compact CSM in both cases. There are no indications of pre-explosion outbursts in either of these CCSNe. This limits the possibility of the compact CSM being modeled with a relatively long-lived CSM. \cite{MoriyaSingh2024} and \cite{ChenTW2024} suggest this CSM be the wind acceleration zone of SN~2023ixf and SN~2024ggi, respectively.  In section \ref{sec:CSM} I concentrated on the recent model by \cite{ChenTW2024} and argued that the acceleration-wind zone model of the CSM requires an unlikely large number of photon scattering events in the radial direction, i.e., $\eta_{\rm w} \gg 1$. Instead, I proposed the effervescent zone model for a pre-explosion dense compact CSM of SN~2024ggi, as I suggested for SN~2023ixf \citep{Soker2023Effer}, or the similar model of \cite{FullerTsuna2024}. I encourage future studies of ejecta-CSM interaction of CCSNe, including SN~2023ixf and SN~2024ggi, not to ignore the effervescent zone model and similar ones (e.g., \citealt{Soker2023Effer, FullerTsuna2024}). 
   
\cite{Singhetal2024} and \cite{ChenTW2024} also find their best fitting to give explosion energy of $E_{\rm exp}  \simeq 2 \times 10^{51} \erg$ for both these CCSNe, and \cite{JacobsonGalanetal2024} adopted an explosion energy of $E_{\rm exp} \simeq 1.2 \times 10^{51} \erg$ for SN~2024ggi. In section \ref{sec:Eexp}, I argued that these energies are a tough challenge to the delayed neutrino explosion mechanism. The energetic challenge adds to the extremely tough challenge of the delayed neutrino mechanism to explain point symmetric CCSN remnants, as was pointed out in 2023-2024 (e.g., \citealt{Soker2024Rev}). I concluded in section \ref{sec:Eexp}  with my view that the JJEM best explains the explosion energy of SN~2024ggi. 
 
The existence of a massive effervescent zone requires strong pulsation and envelope convection (e.g., \cite{Soker2023Effer}, as was shown in detail by \cite{FullerTsuna2024} for a similar CSM model (I use effervescent to include both). The JJEM requires vigorous pre-explosion core convection (e.g., \citealt{PapishSoker2011, GilkisSoker2014, GilkisSoker2016, ShishkinSoker2021, ShishkinSoker2022, WangShishkinSoker2024}). The delayed neutrino explosion mechanism also requires pre-collapse convection to induce perturbations in the collapsing core that aid explosion (e.g., \citealt{Couchetal2015, Muller2020, Bolligetal2021, Yamadaetal2024}). I conclude that the correct modeling of convection, crucially in the inner core but also important in the envelope, is mandatory to explore the explosion and evolution of CCSNe. 

\section*{Acknowledgments}

I thank Luc Dessart  and an anonymous referee for their helpful comments. 
This research was supported by a grant from the Pazy Foundation.


\label{lastpage}

\end{document}